# An Analysis of Thickness-shear Vibrations of an Annular Plate with the Mindlin Plate Equations

Ji Wang, Hui Chen, Tingfeng Ma, Jianke Du, Lijun Yi
School of Mechanical Engineering & Mechanics
Ningbo University, Ningbo, Zhejiang 315211, CHINA
Email: wangji@nbu.edu.cn

Yook-Kong Yong
Department of Civil and Environmental Engineering,
Rutgers University, 623 Bowser Road,
NJ 08854, USA

*Abstract*—The Mindlin plate equations with the consideration of thickness-shear deformation as an independent variable have been used for the analysis of vibrations of quartz crystal resonators of both rectangular and circular types. The Mindlin or Lee plate theories that treat thickness-shear deformation as an independent higher-order vibration mode in a coupled system of two-dimensional variables are the choice of theory for analysis. For circular plates, we derived the Mindlin plate equations in a systematic manner as demonstrated by Mindlin and others and obtained the truncated two-dimensional equations of closely coupled modes in polar coordinates. We simplified the equations for vibration modes in the vicinity of fundamental thickness-shear frequency and validated the equations and method. To explore newer structures of quartz crystal resonators, we utilized the Mindlin plate equations for the analysis of annular plates with fixed inner and free outer edges for frequency spectra. The detailed analysis of vibrations of circular plates for the normalized frequency versus dimensional parameters provide references for optimal selection of parameters based on the principle of strong thickness-shear mode and minimal presence of other modes to enhance energy trapping through maintaining the strong and pure thickness-shear vibrations insensitive to some complication factors such as thermal and initial stresses.

## I. Introduction

High frequency vibrations of plates have been traditionally analyzed with Mindlin [1-5] and Lee plate theories [6-7], which are based on the expansion of displacements and electrical potential of the linear theory of piezoelectricity in power and trigonometric series. Mindlin plate theory, including effects of additional plate rigidity due to shear deformation and plate inertia due to rotations, is the foundation of analysis of quartz crystal resonators, and tremendous efforts have been made to improve the equations and obtain accurate solutions in past decades. For most quartz resonators, shapes of the crystal plates are either rectangular or circular, posing challenges in the analysis with such considerations of complications. The vibrations of rectangular crystal plates based on the Mindlin plate theory have been studied by earlier investigators in different aspects including the analysis of straight-crested waves. From earlier work on vibrations of rectangular crystal plates in Cartesian coordinates with the corrected Mindlin plate equations, we are familiar with the dispersion relations, frequency spectra, and mode shapes, which are playing important roles in the resonator design process. For circular plates, a theoretical analysis of vibrations is more mathematically challenging than rectangular plates and results are also limited.

The Mindlin and Lee plate equations with the consideration of thickness-shear deformation as independent variables have been used for the analysis of vibrations of quartz crystal resonators of both rectangular and circular types. The objective of such analysis is always on identifying optimal parameters of plates and resonator structures to enhance the strong and pure vibrations of the thickness-shear mode, which is the functioning mode of thickness-shear type resonators of both AT- and SC-cut quartz crystal. This goal can only be sufficiently achieved through accurate analysis of vibrations of coupled modes in a finite crystal plates with the consideration of closely clustered modes and complication factors such as electrodes and mounting supports which can affect the frequency and stability under loadings like impact and temperature. The Mindlin or Lee plate theories that treat thickness-shear deformation as an independent higher-order vibration mode in a coupled system of two-dimensional variables are the choice of theory for analysis [8-12].

For circular plates, we had derived the Mindlin plate equations in a systematic manners as demonstrated by Mindlin and others and obtained the truncated two-dimensional equations of closely coupled modes in a finite circular plate. Furthermore, we simplified the equations for modes in the vicinity of fundamental thickness-shear frequency and validated the equations and method from earlier studies [13-14]. In this paper, for the purpose of searching newer structures of quartz crystal resonators, we utilize the first-order Mindlin plate equations for the analysis of annular plates with clamped boundary conditions at the inner edge and free boundary conditions at the outer edge for frequency spectra with the finding that we can obtain similar spectra and vibration modes of a circular plate with free edges for isotropic materials. The radial thickness-shear, flexural, and the circumferential thickness-shear modes are considered in the analysis. The detailed analysis of vibrations of plates for the normalized frequency versus dimensional parameters in the vicinity of

This work was supported by the National Natural Science Foundation of China through grants 11372145 and 11372146.

fundamental thickness-shear frequency will provide guidelines for optimal selection of plate parameters based on the principle of strong thickness-shear mode and minimal presence of other modes to enhance energy trapping through maintaining the strong and pure thickness-shear vibrations insensitive to many complication factors such as thermal and initial stresses.

## II. FIRST-ORDER PLATE EQUATIONS OF THICKNESS-SHEAR VIBRATIONS OF ANNULAR MINDLIN PLATES

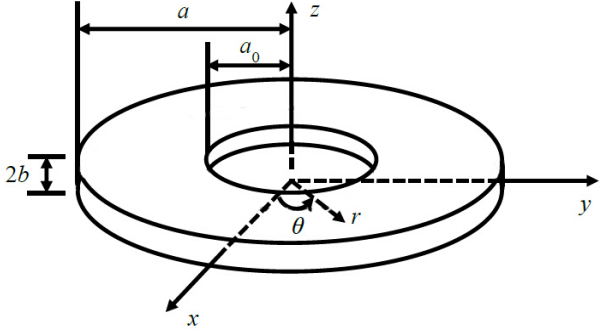

Fig. 1. An annular plate and coordinate system.

We consider the isotropic, annular plate, for which the thickness coordinate is designated by $z$, as shown in Fig. 1. Its inner radius, outer radius, thickness, and mass density are $a_0$, $a$, $2b$ and $\rho$, respectively. With the Mindlin plate equations in polar coordinates, we have the first-order stress equations of motion with coupling of the fundamental thickness-shear $u_r^{(1)}$, flexural $u_z^{(0)}$, and circumferential thickness-shear $u_\theta^{(1)}$ [13-14]

$$T_{rz,r}^{(0)} + \frac{1}{r}T_{\theta z,\theta}^{(0)} + \frac{T_{rz}^{(0)}}{r} = 2\rho b \ddot{u}_z^{(0)},$$
$$T_{rr,r}^{(1)} + \frac{1}{r}T_{\theta r,\theta}^{(1)} + \frac{T_{rr}^{(1)} - T_{\theta\theta}^{(1)}}{r} - T_{zr}^{(0)} = \frac{2b^3}{3}\rho \ddot{u}_r^{(1)}, \quad (1)$$
$$T_{r\theta,r}^{(1)} + \frac{1}{r}T_{\theta\theta,\theta}^{(1)} + 2\frac{T_{r\theta}^{(1)}}{r} - T_{z\theta}^{(0)} = \frac{2b^3}{3}\rho \ddot{u}_\theta^{(1)},$$

where $T_{pq}^{(n)}(p=r,z,\theta;q=r,z,\theta;n=0,1)$ is the $n$ th-order stress.

The constitutive equations for isotropic plates are

$$T_{z\theta}^{(0)} = 2b\mu\kappa^2\left(u_\theta^{(1)} + \frac{1}{r}u_{z,\theta}^{(0)}\right),$$
$$T_{rz}^{(0)} = 2b\mu\kappa^2\left(u_{z,r}^{(0)} + u_r^{(1)}\right),$$
$$T_{rr}^{(1)} = D\left[u_{r,r}^{(1)} + \nu\frac{1}{r}\left(u_{\theta,\theta}^{(1)} + u_r^{(1)}\right)\right], \quad (2)$$
$$T_{\theta\theta}^{(1)} = D\left[\frac{1}{r}\left(u_{\theta,\theta}^{(1)} + u_r^{(1)}\right) + \nu u_{r,r}^{(1)}\right],$$
$$T_{\theta r}^{(1)} = \frac{2}{3}b^3\mu\left[\frac{1}{r}u_{r,\theta}^{(1)} + u_{\theta,r}^{(1)} - \frac{1}{r}u_\theta^{(1)}\right],$$

where $D = 2b^3E/[3(1-\nu^2)]$, and $E$ and $\nu$ are Young's modulus and Poisson's ratio, respectively, and $\kappa^2 = \pi^2/12$. We can also use Lamé constants $\lambda$ and $\mu$ with these equations.

By inserting (2) into (1) and omitting the time factor $\exp(i\omega t)$, the equations of motion for $u_z^{(0)}, u_r^{(1)}$, and $u_\theta^{(1)}$ are

$$\kappa^2\mu\left(\frac{\partial}{\partial r} + \frac{1}{r}\right)u_r^{(1)} + \frac{\kappa^2\mu}{r}u_\theta^{(1)}$$
$$+\left[\kappa^2\mu\left(\frac{\partial^2}{\partial r^2} + \frac{1}{r^2}\frac{\partial^2}{\partial \theta^2} + \frac{1}{r}\frac{\partial}{\partial r}\right) + \rho\omega^2\right]u_z^{(0)} = 0,$$
$$\left[D\frac{\partial^2}{\partial r^2} + \frac{D}{r}\frac{\partial}{\partial r} + \frac{D(1-\nu)}{2r^2}\frac{\partial^2}{\partial \theta^2} - \left(\frac{D}{r^2} + 2\kappa^2\mu b - \frac{2}{3}\rho\omega^2 b^3\right)\right]u_r^{(1)}$$
$$+\left[\frac{D(1+\nu)}{2r}\frac{\partial^2}{\partial \theta \partial r} - \frac{D(3-\nu)}{2r^2}\frac{\partial}{\partial \theta}\right]u_\theta^{(1)} - 2\kappa^2\mu b\frac{\partial u_z^{(0)}}{\partial r} = 0,$$
$$\left[\frac{D(1+\nu)}{2r}\frac{\partial^2}{\partial \theta \partial r} + \frac{D(3-\nu)}{2r^2}\frac{\partial}{\partial \theta}\right]u_r^{(1)}$$
$$+\left[\frac{D(1-\nu)}{2}\frac{\partial^2}{\partial r^2} + \frac{D}{r^2}\frac{\partial^2}{\partial \theta^2} + \frac{D(1-\nu)}{2r}\frac{\partial}{\partial r}\right]u_\theta^{(1)}$$
$$-\left[\frac{D(1-\nu)}{2r^2} + 2\kappa^2\mu b - \frac{2}{3}\rho\omega^2 b^3\right]u_\theta^{(1)} - \frac{2\kappa^2\mu b}{r}\frac{\partial u_z^{(0)}}{\partial \theta} = 0.$$
(3)

Adding (3)$_2$ and (3)$_3$ together yields the governing equations of non-axially symmetric vibrations [10]

$$\frac{b^2}{3}\left[\mu\nabla^2\mathbf{u}_T^{(1)} + \frac{3\lambda\mu+2\mu^2}{\lambda+2\mu}\nabla\left(\nabla\cdot\mathbf{u}_T^{(1)}\right)\right] - \kappa^2\mu\left(\mathbf{u}_T^{(1)} + \nabla u_z^{(0)}\right) = \frac{\rho b^2}{3}\ddot{\mathbf{u}}_T^{(1)},$$
$$\kappa^2\mu\left(\nabla^2 u_z^{(0)} + \nabla\cdot\mathbf{u}_T^{(1)}\right) = \rho\ddot{u}_z^{(0)},$$
(4)

where

$$\mathbf{u}_T^{(1)} = u_r^{(1)}\mathbf{e}_r + u_\theta^{(1)}\mathbf{e}_\theta,$$
$$\nabla = \mathbf{e}_r\frac{\partial}{\partial r} + \mathbf{e}_\theta\frac{1}{r}\frac{\partial}{\partial \theta}. \quad (5)$$

The boundary conditions for fixed inner and free outer edges are

$$u_z^{(0)} = u_r^{(1)} = u_\theta^{(1)} = 0, \text{ at } r = a_0,$$
$$T_{rz}^{(0)} = T_{rr}^{(1)} = T_{r\theta}^{(1)} = 0, \text{ at } r = a. \quad (6)$$

## III. SOLUTIONS OF FREE VIBRATIONS

For non-axially symmetric vibrations of annular plates, we look for waves in the following form

$$u_z^{(0)} = bA_1 J_1(\delta r)\cos\theta + bB_1 Y_1(\delta r)\cos\theta,$$
$$\nabla\cdot\mathbf{u}_T^{(1)} = \frac{1}{b}A_2 J_1(\delta r)\cos\theta + \frac{1}{b}B_2 Y_1(\delta r)\cos\theta, \quad (7)$$

where $A_i(B_i)(i=1,2)$ and $\delta$ are amplitudes and wavenumber, respectively. The normalized variables are utilized as

$$X = \frac{\delta}{\frac{\pi}{2b}}, \Omega = \frac{\omega}{\omega_0}, \omega_0 = \frac{\pi}{2b}\sqrt{\frac{\mu}{\rho}}, \quad (8)$$

with $\omega_0$ as the fundamental thickness-shear vibration frequency.

The substitution of (7) and (8) into (4) yields

$$3\pi^2\kappa^2 X^2 A_1 + \left[\pi^2\Omega^2 - \frac{2\pi^2}{1-\nu}X^2 - 12\kappa^2\right]A_2 = 0,$$
$$\pi^2(\Omega^2 - \kappa^2 X^2)A_1 + 4\kappa^2 A_2 = 0, \quad (9)$$

and

$$3\pi^2\kappa^2 X^2 B_1 + \left[\pi^2\Omega^2 - \frac{2\pi^2}{1-\nu}X^2 - 12\kappa^2\right]B_2 = 0,$$
$$\pi^2(\Omega^2 - \kappa^2 X^2)B_1 + 4\kappa^2 B_2 = 0. \quad (10)$$

For nontrivial solution of $A_1$ and $A_2$, or $B_1$ and $B_2$, the determinant of the coefficient matrix of (9) and (10) must vanish, i.e.,

$$\begin{vmatrix} 3\pi^2\kappa^2 X^2 & \pi^2\Omega^2 - \frac{2\pi^2}{1-\nu}X^2 - 12\kappa^2 \\ \pi^2(\Omega^2 - \kappa^2 X^2) & 4\kappa^2 \end{vmatrix} = 0, \quad (11)$$

Now we need to obtain the components of $\mathbf{u}_T^{(1)}$, which can be achieved by following the procedure developed by Lee et al. [10]. The components of $\mathbf{u}_T^{(1)}$ are

$$u_r^{(1)} = \frac{4}{\pi^2(\Omega^2-1)}\left[-\sum_{i=1}^{2}A_i\beta_i\delta_i b J_1'\left(\frac{\pi X_i}{2}\frac{r}{b}\right) + A_3\frac{b}{r}J_1(\delta_3 r)\right]\cos\theta$$
$$+ \frac{4}{\pi^2(\Omega^2-1)}\left[-\sum_{i=1}^{2}B_i\beta_i\delta_i b Y_1'\left(\frac{\pi X_i}{2}\frac{r}{b}\right) + B_3\frac{b}{r}Y_1(\delta_3 r)\right]\cos\theta,$$
$$u_\theta^{(1)} = \frac{4}{\pi^2(\Omega^2-1)}\left[\frac{b}{r}\sum_{i=1}^{2}A_i\beta_i J_1\left(\frac{\pi X_i}{2}\frac{r}{b}\right) - A_3\delta_3 b J_1'(\delta_3 r)\right]\sin\theta$$
$$+ \frac{4}{\pi^2(\Omega^2-1)}\left[\frac{b}{r}\sum_{i=1}^{2}B_i\beta_i Y_1\left(\frac{\pi X_i}{2}\frac{r}{b}\right) - B_3\delta_3 b Y_1'(\delta_3 r)\right]\sin\theta,$$

(12)

where

$$\beta_i = \frac{8\lambda+8\mu}{\lambda+2\mu} - 3\kappa^2\alpha_i, \quad \alpha_i = -\frac{4\kappa^2}{\pi^2(\Omega^2-\kappa^2 X^2)}, i=1,2,$$
$$\delta_3^2 = \frac{\pi^2}{4b^2}\eta, \quad \eta = \Omega^2 - 1.$$

(13)

A substitution of stresses and displacements into the six boundary conditions at $r = a_0$, and $r = a$ in (6) yields to six linear equations for the six amplitudes $A_i$ $(B_i)(i = 1,2,3)$. As usual, by setting the coefficient determinant of the boundary condition equations to vanish, frequency spectra and modes shapes can be calculated as the functions of parameters of the plate dimensions and material properties, as shown in equations of stresses and displacements. As usual, the frequency equation is given in transcendental functions and the solutions of roots or zero-points require accurate evaluation of Bessel functions. The algorithms for the numerical procedures are generally known and can be implemented with mathematical tools.

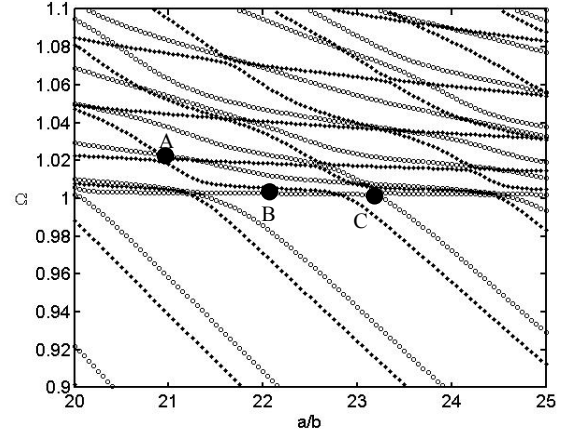

Fig. 2. Frequency spectra of a free circular plate (•) vs. an annular plate (○) with the Mindlin plate equations.

For clamped edge conditions at the inner ($a_0 = 2b$) and free edge conditions at the outer boundaries of the annular plate, we obtained frequency spectra similar to a free circular plate as shown in Fig. 2. In order to identify the frequency features at the specific frequency at which the three vibration modes are coupled, we calculated mode shapes at various locations of the spectra in Fig. 2 labeled with A, B, and C. Figs. 3-5 shows the distribution of relative displacements at frequencies corresponding to points A, B, and C. The results show that the vibration modes depend on the frequency strongly. At point A, there is a strong coupling between all the vibration modes and the flexure $u_z^{(0)}$ is strong. At point B, we can guess that it is the dominant thickness-shear mode $u_r^{(1)}$, which has relatively large amplitude. Similarly, point C is another location with strong couplings of three modes and the transverse thickness-shear mode $u_\theta^{(1)}$ is relatively strong. For resonator design, we always want to find the strong radial thickness-shear mode with less coupling to the flexural mode to achieve better resonance with less energy dissipation. The frequency spectra in Fig. 2 and the mode shapes in Figs. 3-5 are the needed information to find the right parameters and modes.

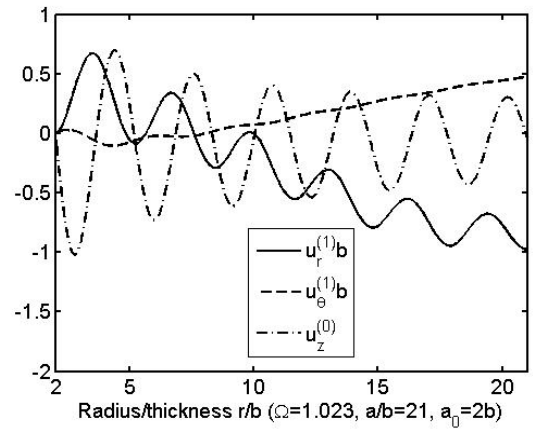

Fig. 3. Distribution of displacements at frequency corresponding to point A.

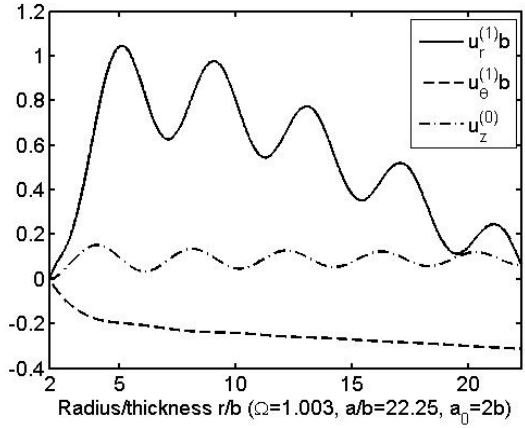

Fig. 4. Distribution of displacements at frequency corresponding to point B.

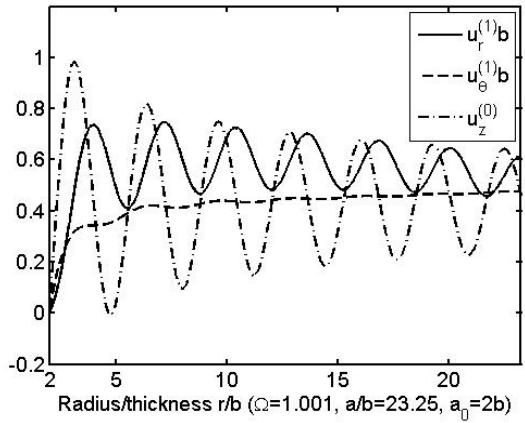

Fig. 5. Distribution of displacements at frequency corresponding to point C.

## IV. CONCLUSIONS

We had utilized the first-order Mindlin plate equations for the analysis of the fundamental thickness-shear vibrations of an isotropic, annular plate, and successfully solved the coupled equations with the Bessel functions in a manner similar to the trigonometric functions in the analysis of vibrations of rectangular plates. With these validated equations and solutions experiences, we obtained frequency spectra and modes shapes of an annular plate with clamped boundary conditions at the inner edge and free boundary conditions at the outer edge. The results showed that there are frequencies reflecting dominant vibration modes and their strength of couplings. We can use the spectra to find ideal regions with the dominant thickness-shear vibrations for possible designs of quartz crystal resonators with different structures. The procedure outlined in this study can be used for the further search of novel configuration of quartz crystal resonators with new materials, structures, and features.

REFERENCES

[1] R.D. Mindlin, "Influence of rotatory inertia and shear on flexural vibrations of isotropic, elastic plates," J. Appl. Mech., vol. 18, 1951, pp. 31-38.

[2] R.D. Mindlin, "Forced thickness-shear and flexural vibrations of piezoelectric crystal plates," J. Appl. Phys., vol. 23, 1952, pp. 83-88.

[3] H.F. Tiersten and R.D. Mindlin, "Forced vibrations of piezoelectric crystal plates," Q. Appl. Math., vol. 20, 1962, pp. 107-119.

[4] R.D. Mindlin, "High frequency vibrations of piezoelectric crystal plates," Intl. J. Solids Struct., vol. 8, 1972, pp. 895-906.

[5] R.D. Mindlin (edited by J.S. Yang), An Introduction to the Mathematical Theory of Vibrations of Elastic Plates, World Scientific, New Jersey, 2007.

[6] P.C.Y. Lee and Z. Nikodem, "An approximate theory for high frequency vibrations of elastic plates," Intl. J. Solids Struct., vol. 8, 1972, pp. 581-611.

[7] P.C.Y. Lee, J.D. Yu, and W.S. Lin, "A new two-dimensional theory for vibrations of piezoelectric crystal plates with electroded faces," J. Appl. Phys., vol. 83, 1998, pp. 1213-1223.

[8] R.D. Mindlin and H. Deresiewicz, "Thickness-shear and flexural vibrations of a circular disk," J. Appl. Mech., vol. 25, 1954, pp. 1329-1332.

[9] H. Deresiewicz and R.D. Mindlin, "Axially symmetric flexural vibrations of a circular disk," J. Appl. Mech., vol. 22, 1955, pp. 86-88.

[10] P.C.Y. Lee, J.D. Yu, X. Li, and W.-H. Shih, "Piezoelectric ceramic disks with thickness-graded material properties," IEEE Trans. Ultrason., Ferroelect., Freq. Contr., vol. 46, 1999, pp. 205-215.

[11] P.C.Y. Lee, R. Huang, X. Li, and W.-H. Shih, "Vibrations and static responses of asymmetric bimorph disks of piezoelectric ceramics," IEEE Trans. Ultrason., Ferroelect., Freq. Contr., vol. 47, 2000, pp. 706-715.

[12] R. Huang, P.C.Y. Lee, W.S. Lin, and J.D. Yu, "Extensional, thickness-stretch and symmetric thickness-shear vibrations of piezoceramic disks," vol. 49, 2002, pp. 1507-1515.

[13] W.J. Wang, J. Wang, G.J. Chen, T.F. Ma, and J.K. Du, "Mindlin plate equations for the thickness-shear vibrations of circular elastic plates," in Proc. 2012 Symp. on Piezoeletricity, Acoustic Waves, and Device Applications, Shanghai, China, Nov. 23 –25, 2012, pp. 357-360.

[14] H. Chen, J. Wang, T.F. Ma, and J.K. Du, "An analysis of free vibrations of isotropic, circular plates with the Mindlin plate theory," in Proc. 2014 Symp. on Piezoeletricity, Acoustic Waves, and Device Applications, Beijing, China, Oct. 30 – Nov. 2, 2014, pp. 397-400.